\documentclass[fleqn,10pt]{wlscirep}
\usepackage[utf8]{inputenc}
\usepackage[T1]{fontenc}
\title{Diffusion-driven lensless fiber endomicroscopic quantitative phase imaging towards digital pathology}

\author[1,2]{Zhaoqing Chen}
\author[3,*]{Jiawei Sun}
\author[3]{Xibin Yang}
\author[2]{Xinyi Ye}
\author[2,4]{Bin Zhao}
\author[2,5]{Xuelong Li}
\author[6,7]{Juergen W. Czarske}

\affil[1]{Shanghai Research Institute for Intelligent Autonomous Systems, Tongji University, Shanghai, China}
\affil[2]{Shanghai Artificial Intelligence Laboratory, Longwen Road 129, Xuhui District, 200232 Shanghai, China}
\affil[3]{Suzhou Institute of Biomedical Engineering and Technology, Chinese Academy of Sciences, 215163 Suzhou, China}
\affil[4]{School of Artificial Intelligence, OPtics and ElectroNics (iOPEN), Northwestern Polytechnical University, 711072 Xi'an, Shaanxi, China}
\affil[5]{Institute of Artificial Intelligence (TeleAI), China Telecom Corp Ltd, 100033 Beijing, China}
\affil[6]{Competence Center for Biomedical Computational Laser Systems (BIOLAS), TU Dresden, Helmholtzstrasse 18, 01069 Dresden, Germany}
\affil[7]{Laboratory of Measurement and Sensor System Technique (MST), TU Dresden, Helmholtzstrasse 18, 01069 Dresden, Germany}

\affil[*]{sunjiawei@sibet.ac.cn}

\begin{abstract}
Lensless fiber endomicroscopy is an emerging tool for minimally invasive in vivo imaging, where quantitative phase imaging can be utilized as a label-free modality to enhance image contrast. Nevertheless, current phase reconstruction techniques in lensless multi-core fiber endomicroscopy are effective for simple structures but face significant challenges with complex tissue imaging, thereby restricting their clinical applicability. We present SpecDiffusion, a speckle-conditioned diffusion model tailored to achieve high-resolution and accurate reconstruction of complex phase images in lensless fiber endomicroscopy. Through an iterative refinement of speckle data, SpecDiffusion effectively reconstructs structural details, enabling high-resolution and high-fidelity quantitative phase imaging. This capability is particularly advantageous for digital pathology applications, such as cell segmentation, where precise and reliable imaging is essential for accurate cancer diagnosis and classification. New perspectives are opened for early and accurate cancer detection using minimal invasive endomicroscopy.

\end{abstract}
\begin{document}

\flushbottom
\maketitle
%
%
\thispagestyle{empty}


\section*{Introduction}

Lensless fiber endomicroscopy is an emerging tool in medical imaging and diagnostics, noted for its high imaging resolution and minimum invasiveness~\cite{porat2016widefield,borhani2018learning,kuschmierz2021ultra,sun2024ai,sun2022quantitative,badt2022real, choi2022flexible, wang2024resolution, sun2024lensless}. Unlike traditional endoscopes
that rely on lenses of the application side to capture images, lensless fiber endomicroscope employs computational imaging techniques to visualize microscopic structures with an ultra-thin probe. Therefore, it provides access for previously inaccessible area and reduces neuronal damage during clinical operation, thereby minimizing patient discomfort and contributing to faster post-operation recovery~\cite{li2021memory, koukourakis2022investigation, lich2024single}. For typical lensless fiber endomicroscopes employing wavefront shaping technologies, in vivo confocal fluorescent imaging is usually utilized~\cite{schurmann2018three,sun2021complex,flusberg2005fiber,szabo2014spatially, haufe2017transmission, rothe2019transmission}. However, the potential toxin of fluorescent dyes has prompted the exploration of safer alternatives~\cite{flusberg2005fiber}, and approaches are being sought to save the effort and time required for staining tissue~\cite{abraham2023label,park2023artificial,pillar2024virtual}. Quantitative phase imaging (QPI) serves as a promising label-free technique, which utilizes the phase information to enhance endomicroscopic imaging capabilities~\cite{sun2022quantitative, rothe2021benchmarking, koukourakis2022investigation}. Moreover, it enables the extraction of critical biophysical properties of biological samples, such as refractive index~\cite{wang2011tissue, Liu2016}, cell volume~\cite{park2018quantitative} and dry mass~\cite{schurmann2016cell, aknoun2015living}, which has been proven to be a potential biomarker for cancer diagnosis~\cite{parvin2021differential}. Therefore, achieving QPI through lensless fiber endomicroscopes could help cancer diagnosis in the early stage with minimum invasiveness. Nevertheless, due to the optical path difference among different fiber cores of the multi-core fiber (MCF), the phase information gets distorted through the lensless endomicroscope~\cite{kuschmierz2018}. To address this problem, the far-field amplitude-only speckle transfer (FAST) technique~\cite{sun2022quantitative,sun2023compressive} is introduced to correct phase distortion via an additional reference measurement, enabling high-fidelity QPI through the lensless fiber endomicroscope. However, the iterative propagation process and the requirement of additional calibration procedure limit FAST's imaging speed. This constraint significantly hamper its application in real-time medical diagnostics during clinical procedures, thereby hindering the precision of a doctor's clinical operation.

Recently, deep learning emerges as a powerful tool in QPI~\cite{wang2020phase, bostan2020deep, zhou2024deep,liu2024learning, wu2021high, zhang2022learning}. It first verifies its effectiveness in reconstructing phase and amplitude image through a multi-mode fiber~\cite{rahmani2018multimode, borhani2018learning, chen2023deep}, successfully reconstructing simple structures like the handwritten digits in MNIST dataset~\cite{lecun1998gradient}. Nevertheless, multi-mode fiber is sensitive to fiber conformation and polarization, hindering its ability to achieve high-resolution phase imaging. In contrast, MCFs are relatively less susceptible to external factors. Although deep learning has proved its potential in phase reconstruction through MCFs~\cite{sun2024calibration}, the capability of the previously reported methods are limited in reconstructing simple structures like handwritten digits and icons~\cite{deng2012mnist, xiao2017fashion}. The performance of deep learning models is contingent upon the complexity of the images being reconstructed. Consequently, the relatively simplistic architecture of the conventional U-Net ~\cite{ronneberger2015u}, which is generally utilized in previous works, is inadequate for the phase reconstruction of complex images through scattering medium like MCFs. Diffusion model~\cite{song2019generative, ho2020denoising,rombach2022high} is the recently prevailing AI model noted for generating high-quality, complex images. Unlike traditional image-to-image reconstruction methods, the generation procedure of diffusion models involves two key processes: the forward process, where data is gradually transformed into Gaussian noise through a sequence of steps, and the reverse process, where the model learns to reconstruct the original data from this noise. The effectiveness of diffusion model origins from its iterative generation process, which alleviates the generation challenge in each step and useful for tasks where fine details are critical. Hence, diffusion models have been widely implemented in sophisticated tasks including drug design~\cite{igashov2024equivariant}, protein design~\cite{guo2024diffusion} and microscopic image data analysis~\cite{kreis2022latent, waibel2023diffusion, wang2024conditional, guo2024diffusion, wang2024decoding}. While typical diffusion models offer powerful capabilities in unsupervised image generation, applying them in complex phase reconstruction tasks remains challenging. Typically, the unsupervised design of diffusion model hinders phase reconstruction from the speckle images captured at the detection system. The most recent work incorporates physical process into denoising process of the diffusion model, achieving high-quality holographic reconstruction~\cite{zhang2024single}. However, it relies on specific initialization and well-crafted stepsize to avoid undesired artifacts, which undermines the fidelity of the reconstructed images. Therefore, high-fidelity phase reconstruction of complex images using diffusion models remains challenging.

\begin{figure}[t]
\centering
\includegraphics[width=\textwidth]{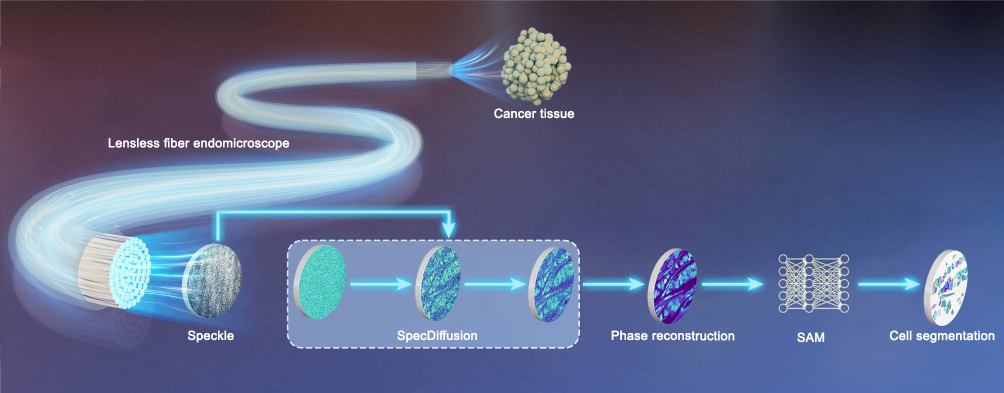}
\caption{Illustration demonstrates the working principle of the diffusion-driven lensless fiber endomicroscope. The quantitative phase reconstruction process involves using speckle images captured at the detection system to guide the denoising process of the SpecDiffusion model. The reconstructed phase images are applicable to digital pathology tasks, including cell segmentation, enhancing both accuracy and detail in diagnostic imaging.}
\label{fig:com}
\end{figure}

In this paper, we introduce a conditional diffusion model, termed SpecDiffusion, to achieve high-fidelity quantitative phase reconstruction in the lensless fiber endomicroscope using speckle images captured at the detection system. As illustrated in Fig.~\ref{fig:com}, SpecDiffusion implements an iterative framework to break the challenging speckle reconstruction task into several steps. Inspired by the phase retrieval algorithms~\cite{gerchberg1972holography}, we use the intensity-only speckle images as a supervisor for the diffusion model in each step to ensure high fidelity phase reconstruction. For the phase retrieval process, SpecDiffusion needs only one intensity image as the reference, and the quality of reconstructed phase image is significantly enhanced through the iterative reconstruction process. To verify the effectiveness of the proposed diffusion model, we build an optical setup to generate training dataset consisting images from ImageNet~\cite{deng2009imagenet} and their corresponding speckles through MCF. Given the complexity and diversity of ImageNet images, this dataset can serve as a benchmark for assessing the phase reconstruction capability of the lensless fiber endomicroscope in various scenarios. The pretrained SpecDiffusion model on this dataset demonstrates high robustness in unseen structures. This is validated by a USAF resolution test chart, where the proposed diffusion-driven endomicroscope demonstrates its ability to resolve details down to several micrometers. Additionally, we explore the practical potential of SpecDiffusion in biomedical applications, achieving cancer tissue image reconstruction with high resolution by transfer learning. To further validate the efficacy of SpecDiffusion's reconstructed tissue images in digital pathology, a zero-shot cell segmentation task is conducted on these reconstructed tissue images. Notably, SpecDiffusion not only achieves improvements in key image quality metrics, but also closely mirrors the segmentation results obtained from the real phase image. Our method offers promising capabilities for phase reconstruction in complex images captured through lensless MCF endomicroscopes, paving the way for advanced applications in cancer diagnosis and metabolic studies.

\section*{Results}


\subsection*{Diffusion-driven quantitative phase imaging through a lensless fiber endomicroscope}

\begin{figure}[ht]
\centering
\includegraphics[width=\textwidth]{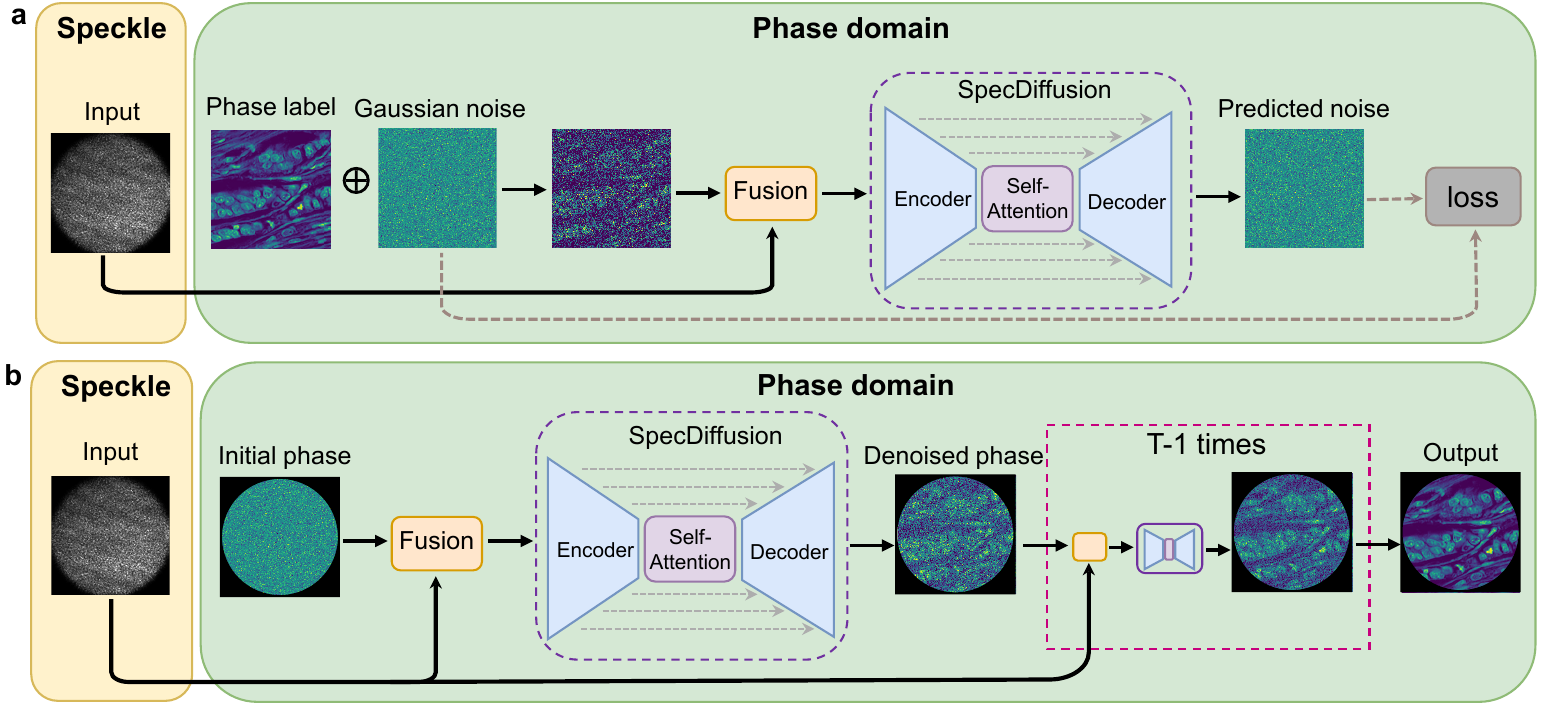}
\caption{Architecture of the speckle-conditioned diffusion model (SpecDiffusion). \textbf{a} Training process of SpecDiffusion. Phase label is mixed with Gaussian noise, and SpecDiffusion is trained to predict the imposed noise with the guidance of speckle. \textbf{b} Inference process of SpecDiffusion. With the guidance from input speckle, the randomly-generated initial phase is gradually denoised and transformed towards label phase by SpecDiffusion.}
\label{fig:model}
\end{figure}

We propose a conditional diffusion model, termed SpecDiffusion, for phase reconstruction of complex images. It breaks down reconstruction process into iterative denoising steps, gradually transforming the initial random phase into desired phase image. The speckle images captured at the far-field of the MCF are served as the conditioning images that guide the iterative denoising process during phase reconstruction. SpecDiffusion is trained from scratch using a supervised approach, ensuring high reconstruction fidelity. As shown in Fig.~\ref{fig:model}a, during training, Gaussian noise is incrementally added to the ground truth phase images over several time steps, degrading them to an almost unrecognizable state. SpecDiffusion then operates as a denoising model, predicting and progressively removing the noise. Unlike conventional diffusion models, each denoising step is guided by conditioning speckle images, steering the noise reduction process in a physics-informed manner and enhancing the accuracy of phase reconstruction.

The inference process employed by the SpecDiffusion is depicted in Fig.~\ref{fig:model}b. In this process, the model learns to reconstruct the phase information incident the MCF from the noisy random initial phase by performing a series of learned denoising steps. Each denoising step utilizes not only the current state of the reconstructed phase but also the speckle images to reverse the noise addition accurately. Subsequently, this denoised phase is mixed with a smaller amount of Gaussian noise, serving as the initialization for the next iteration. This denoising step is repeated for $T$ iterations, progressively reconstructing the phase image towards the actual phase. The culmination of this process yields the final phase image, which accurately represents the predicted phase corresponding to the input speckle pattern. Detailed principles of the SpecDiffusion is explained in Methods section.

\begin{figure}[t]
\centering
\includegraphics[width=\textwidth]{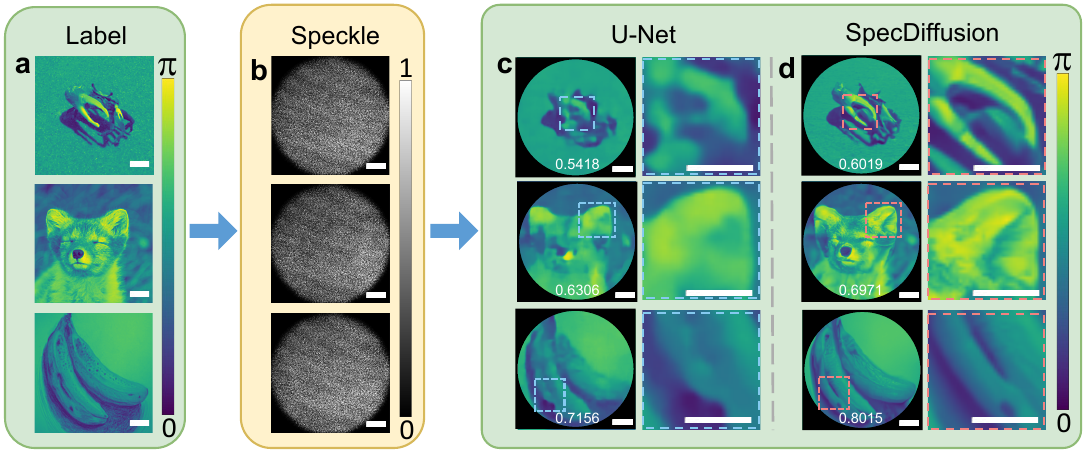}
\caption{Diffusion-driven phase reconstruction of ImageNet images through lensless fiber endomicroscope. \textbf{a} Ground truth phase images. \textbf{b} Speckle patterns captured at the detection side of the lensless fiber endomicroscope. \textbf{c} and \textbf{d} Reconstructed phase images by U-Net and SpecDiffusion. The SSIM value for each reconstructed image with respect to its corresponding phase label is shown. Scale bars $50~ \mu m$.}
\label{fig:imagenet}
\end{figure}

To visualize the phase reconstruction performance of SpecDiffusion, ground truth phase images are illustrated in Fig.~\ref{fig:imagenet}a. They are holographically projected on the facet of MCF. When the modulated light field transmitted through the MCF, its phase gets distorted due to intrinsic optical path differences between cores and the honeycomb artifacts~\cite{wang2024resolution}, thus forming the speckle patterns in Fig.~\ref{fig:imagenet}b. Based on the speckle patterns, we compare the performance of the U-Net and SpecDiffusion in Fig.~\ref{fig:imagenet}c and d. Benefiting from the iterative reconstruction process,  SpecDiffusion demonstrates the capacity to reconstruct detailed structures within complex images. In the magnified region of the crab image, U-Net manages to reconstruct only the approximate shape of the pincers. In contrast, SpecDiffusion achieves a much clearer and more detailed reconstruction of the pincers. Further detailed comparison of the phase  reconstruction results are demonstrated in supplementary Fig.~S$2$. To further quantitatively characterize the phase reconstruction fidelity, we employ the mean absolute metric (MAE), peak signal-to-noise (PSNR), structural similarity index measure (SSIM)~\cite{wang2004image}, 2D correlation coefficient and phase residual standard deviation (PRSTD) as evaluation metrics. The metric distributions of U-Net and SpecDiffusion are illustrated in supplementary Fig.~S$2$e-h. The averaged evaluation metrics are summarized in Table~S$1$. Quantitatively, SpecDiffusion outperforms conventional U-Net on all metrics, indicating higher fidelity of the reconstructed phase images. The evaluation on ImageNet underlines SpecDiffusion's ability to enhance diagnostic imaging by providing clearer, more detailed reconstructions. In medical application, clearer reconstructions result in sharper boundaries between different regions, thereby improving discrimination for cancer diagnosis. Moreover, the detailed reconstructions provide more imaging information retrieved from MCF speckles, which is beneficial for further virtual staining.

\subsection*{Characterization of the resolution through the diffusion-driven fiber endomicroscope}

\begin{figure}[t]
\centering
\includegraphics[width=\textwidth]{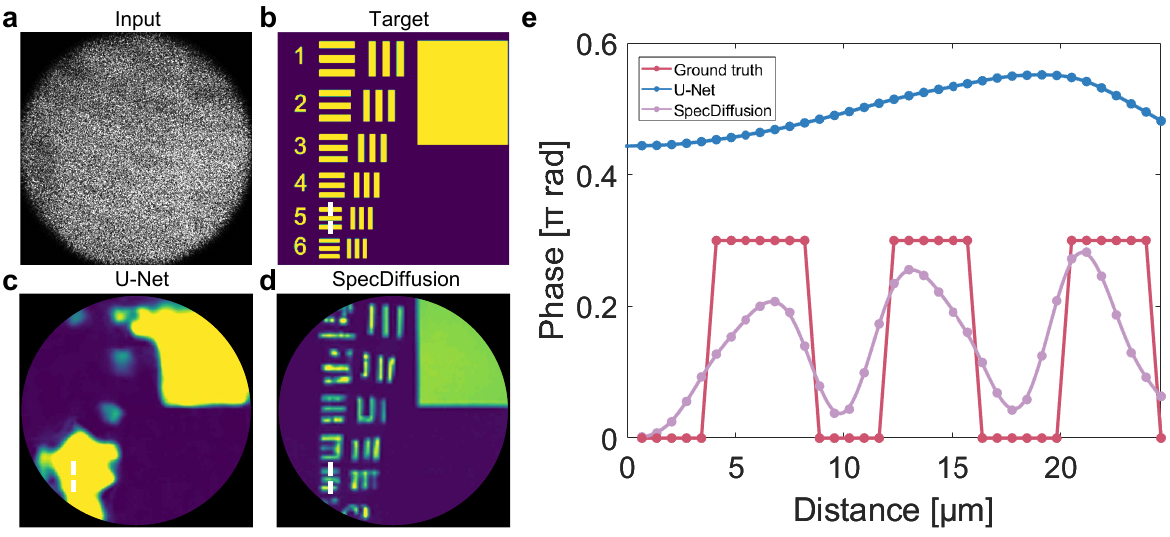}
\caption{Diffusion-driven phase reconstruction of test chart through the lensless fiber endomicroscope. \textbf{a} Speckle pattern captured at the detection side of lensless fiber endomicroscope . \textbf{b} Ground truth images of the test chart. \textbf{c} U-Net reconstructed phase image. \textbf{d} SpecDiffusion reconstructed phase image. \textbf{e} Phase reconstruction contrast of ground truth, U-Net and SpecDiffusion. Scale bars $50~ \mu m$.}
\label{fig:res}
\end{figure}

The evaluation of natural images is crucial for evaluating the general performance of SpecDiffusion. However, to precisely quantify the resolution of the diffusion-driven fiber endomicroscope, it is essential to employ the standardized USAF 1951 test chart as the evaluation target. Reconstructing the test chart using a deep neural network presents a unique challenge because it is an unseen object with distinct high-frequency features that differ significantly from the natural images used to train such models. Deep neural networks, particularly those trained on natural image datasets, are often biased toward capturing low-frequency, global features, which may result in poor reconstruction of fine, high-frequency details like those found in the test chart~\cite{durall2020watch}. This issue is exacerbated in speckle-based imaging systems, where the random and complex nature of speckle patterns further complicates the process of accurately resolving small, detailed structures. As a result, the USAF test chart is a widely recognized benchmark for evaluating the performance of neural networks in resolving fine details, especially when applied to unseen objects with features that differ from the training data.

In the experiment, phase images of the USAF chart were projected onto the spatial light modulator (SLM), with the corresponding speckle patterns captured by the detection system. The reconstructed images were then analyzed to determine the system's resolution limit by identifying the smallest resolvable elements on the test chart. As depicted in Fig.~\ref{fig:res}d, the results demonstrate that the diffusion-driven fiber endomicroscope is capable of resolving bars with a width of $4.10~\mu m$, indicating a precision suitable for complex imaging tasks. In comparison, conventional neural networks like U-Net can hardly achieve high resolution reconstruction in this range. As demonstrated in Fig.~S$3$, the optimum resolution of the system using the U-Net for phase reconstruction is $10.25~\mu m$. In Fig.~\ref{fig:res}e, the reconstruction results from SpecDiffusion effectively delineate the peak phase across three lines, exhibiting a pronounced contrast between the peaks and valleys within the phase image. This distinct clarity demonstrates that SpecDiffusion, through the accurate reconstruction of speckle patterns, enables micrometer-level imaging resolution for the lensless fiber endomicroscope. 
This reveals that the diffusion-driven fiber endomicroscope can be a promising high-resolution imaging probe for biomedical applications.



\subsection*{Diffusion-driven fiber-optic phase reconstruction towards human cancer tissue samples}
\begin{figure}[t!]
\centering
\includegraphics[width=\textwidth]{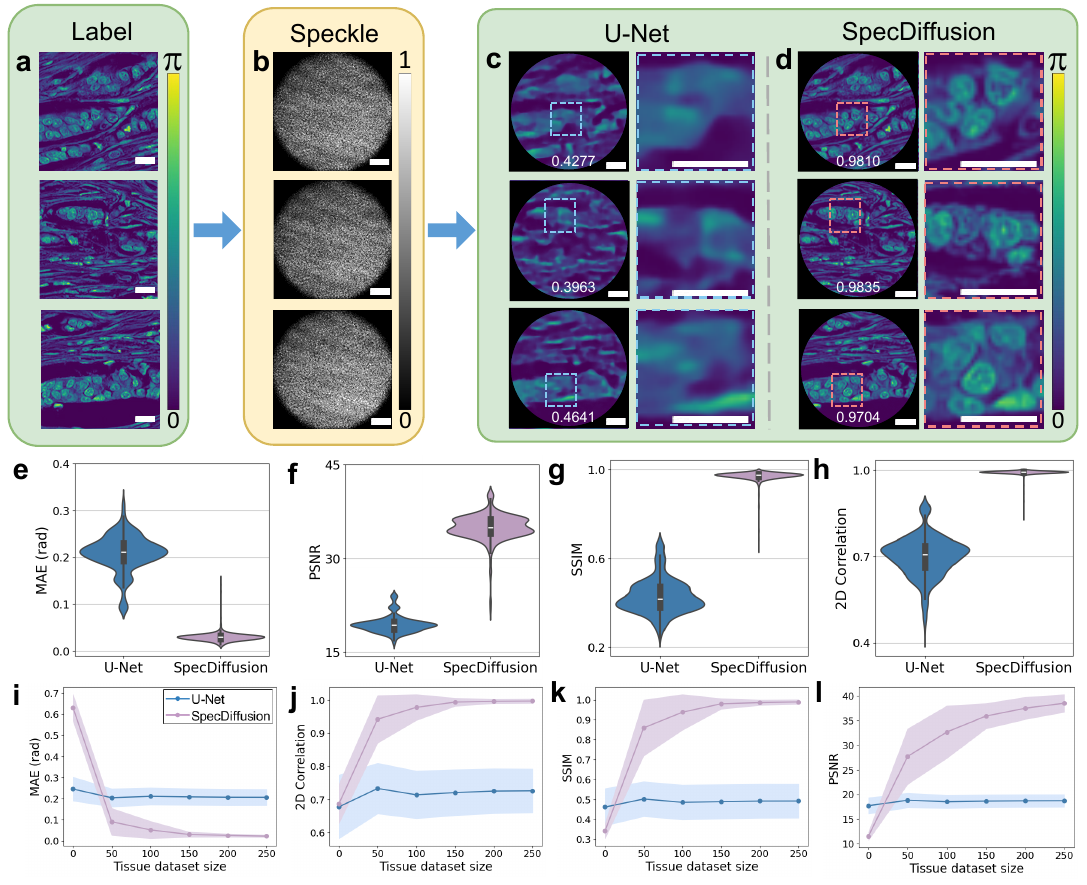}
\caption{Diffusion-driven cancer tissue reconstruction through the lensless fiber endomicroscope. \textbf{a} Ground truth phase images. \textbf{b} Speckle patterns from MCF captured at the detection side of the lensless fiber endomicroscope. \textbf{c} and \textbf{d} Reconstructed phase image by U-Net and SpecDiffusion. \textbf{e} MAE, \textbf{f} PSNR, \textbf{g} SSIM and \textbf{h} 2D correlation coefficient distribution evaluated on $1,000$ reconstructed tissue images by U-Net and SpecDiffusion. \textbf{i} MAE, \textbf{j} PSNR, \textbf{k} SSIM and \textbf{l} 2D correlation coefficient evaluated on $1,000$ reconstructed tissue images by U-Net and SpecDiffusion with varying tissue dataset size. The SSIM value for each reconstructed image with respect to its corresponding phase label is shown. Scale bars $50~ \mu m$.}
\label{fig:cell}
\end{figure}

To further evaluate SpecDiffusion's potential in biomedical applications, such as cancer diagnosis, we demonstrate an additional dataset consisting of human cancer tissue. Typically, deep learning models encounter significant challenges with overfitting, resulting in difficulties when generalizing across various application domains, particularly in the medical field. Nevertheless, the diversity inherent in the ImageNet dataset, combined with SpecDiffusion's robust reconstruction abilities, fortifies our model against overfitting. Notably, with transfer learning on a small set of tissue, which is readily obtainable in clinical scenarios, SpecDiffusion not only sustains its performance but also exhibits improved efficacy. A reconstructed video demonstrating the denoising process of SpecDiffusion is shown in Supplementary Video $1$. 

Based on the model pretrained on ImageNet, we finetune it on $150$ tissue images, and the reconstructed images of U-Net and SpecDiffusion are depicted in Fig.~\ref{fig:cell}c and d. SpecDiffusion clearly outlines the cell morphology and boundaries, facilitating precise cell analysis and enumeration which are crucial for b. Moreover, it vividly captures cellular sociology, depicting the cellular aggregation patterns within tissues. The MAE, PSNR, SSIM and 2D correlation coefficient distributions of U-Net and SpecDiffusion are illustrated in Fig.~\ref{fig:cell}e-h. The distributions of SpecDiffusion are concentrated in regions indicative of better metric performance and exhibit a more centered tendency on most metrics. This suggests that the phase reconstruction capacity of SpecDiffusion is reliable and consistent. The averaged evaluation metrics are summarized in Table~\ref{tbl:cell}. Quantitatively, SpecDiffusion reduces MAE from $0.2091~ rad$ to $0.0304~ rad$, indicating its accurate reconstruction. This reduction also highlights SpecDiffusion's potential for further applications in measuring biophysical properties, such as refractive index, dry mass and optical transmission matrix~\cite{koukourakis2022investigation}. As a result, real-time quantitative measurement during clinical operation becomes possible, enabling precise identification of boundaries between healthy and diseased regions, and minimizing patient harm. Meanwhile, PSNR is improved from $19.43~dB$ to $34.98~dB$, suggesting that SpecDiffusion extracts more information from speckles, thereby supporting the success of subsequent virtual staining. SpecDiffusion also achieves higher SSIM from $0.4310$ to $0.9697$, which is an indicator for retaining more structural information, making the reconstructed tissue structures reliable for clinical cancer diagnosis. Furthermore, the elevated 2D correlation coefficient, from $0.6999$ to $0.9915$, and the improved PRSTD value, from $0.2532$ to $0.0422$, demonstrate greater overall fidelity with less reconstruction bias between different regions, further proving the reliability of the tissue images. These metrics collectively affirm that, SpecDiffusion's reconstructions are high-fidelity and reliable, thereby demonstrating its applicability to complex image reconstruction tasks.  More tissue reconstruction results and the corresponding residual maps are illustrated in supplementary Fig.~S$4$. The residual maps demonstrate low reconstruction errors for SpecDiffusion, verifying its phase reconstruction precision. This enhanced performance and precision makes SpecDiffusion a promising tool for advanced medical imaging and diagnostics, offering significant benefits for clinical practices.

\begin{figure}[t]
\centering
\includegraphics[width=\textwidth]{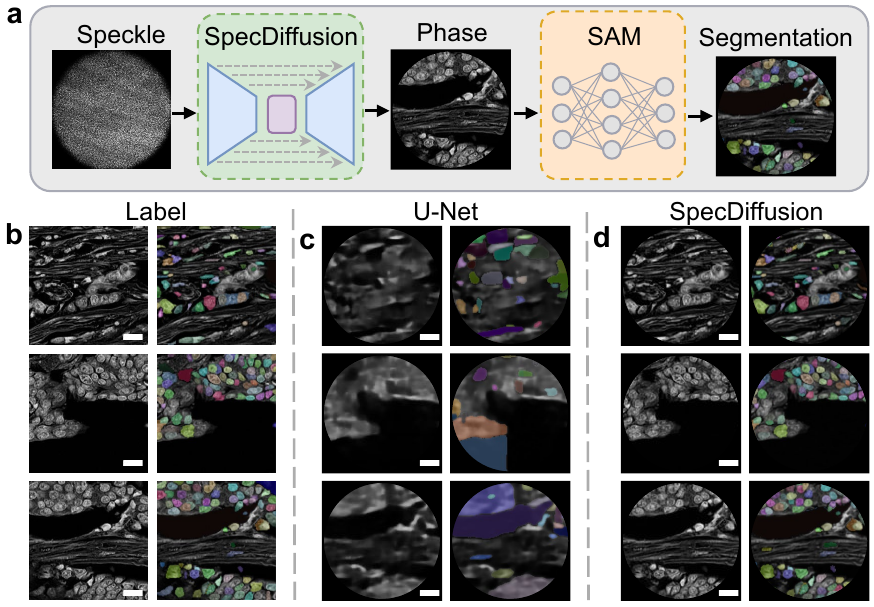}
\caption{Cell segmentation by SAM on ground truth and reconstructed images. \textbf{a} Cell segmentation based on lensless MCF phase imaging by SAM. \textbf{b} Ground truth phase images and their cell segmentation results. \textbf{c} and \textbf{d} Reconstructed phase images by U-Net and SpecDiffusion, and their cell segmentation results. Scale bars $50~ \mu m$}
\label{fig:seg}
\end{figure}

Additionally, we examine how the size of the tissue dataset used for transfer learning impacts the reconstruction capabilities of SpecDiffusion, as detailed in Fig.~\ref{fig:cell}i-l. Initially, SpecDiffusion displays limited reconstruction efficiency when applied directly to the tissue dataset. However, it rapidly adapts to the tissue reconstruction task and improves its performance with the support of modestly-sized dataset. In contrast, the conventional U-Net model shows only marginal enhancements in its reconstruction abilities through transfer learning, remaining substantially less effective overall. Moreover, we investigated how the number of denoising steps in SpecDiffusion influences its reconstruction quality, as depicted in supplementary Fig.~S$5$. By increasing the denoise steps, SpecDiffusion sacrifices inference speed for greater reconstruction accuracy and fidelity. At a denoise step number of $100$, SpecDiffusion achieves a reconstruction speed of $1.77$ frame per second (FPS). It renders SpecDiffusion a feasible tool for clinical applications where time sensitivity is not a critical concern.

To further assess the potential of our reconstructed images in conjunction with other AI technologies, we employ the segmentation anything model (SAM)~\cite{kirillov2023segment} for zero-shot cell segmentation tasks on reconstructed and ground truth images, as illustrated in Fig.~\ref{fig:seg}a. SAM stands out as an AI segmentation model characterized by its exceptional generalization ability, capable of delivering outstanding segmentation outcomes without the prior knowledge about the samples. The details for segmentation task is explained in Methods section. As AI technologies like SAM are increasingly integrated into medical diagnostics, it's important to ensure that the reconstructed images maintain fidelity to their originals within the realm of modern AI technology.

\begin{table}[htb]
\centering
  \caption{Phase reconstruction through lensless fiber endomicroscope using U-Net and SpecDiffusion on human cancer tissue dataset}
  \label{tbl:cell}
  \begin{tabular}{ccccccc}
    \hline
    Method  & MAE(rad) & PSNR & SSIM & 2D Correlation & PRSTD  & IoU\\
    \hline
    U-Net   & 0.2091  & 19.43 & 0.4310 & 0.6999 & 0.2532  & 0.2480\\
    SpecDiffusion   & 0.0304 & 34.98  &0.9697 & 0.9915 & 0.0422  & 0.7658\\
    \hline
  \end{tabular}
\end{table}

The segmentation results of SpecDiffusion's reconstructed images are illustrated in Fig.~\ref{fig:seg}d, which closely mirror those of the ground truth in Fig.~\ref{fig:seg}b, accurately depicting cell locations, sizes and morphologies. This performance starkly contrasts with U-Net, whose segmentation results, indicated in Fig.~\ref{fig:seg}c, are limited to identifying the general boundaries of different tissues without providing the granularity required for detailed cell-level segmentation. To quantitatively evaluate the segmentation quality, we employ the intersection over union (IoU) to measure the agreement between segmentation results derived from reconstructed phase images and their ground truth counterparts. The methodology for calculating IoU is detailed in Methods section. The averaged IoU is illustrated in Table~\ref{tbl:cell}. SpecDiffusion substantially enhances segmentation accuracy, improving the IoU from $0.2480$ to $0.7658$. The distribution of IoU values, highlighted in supplementary Fig.~S$6$, indicates the aggregation of SpecDiffusion’s segmentation results within the higher IoU range, suggesting that the majority of its reconstructions are conducive to effective segmentation. The results signify the alignment between SpecDiffusion's reconstructed images and the ground truth in the context of segmentation tasks, indicating its higher fidelity for downstream applications.

\section*{Discussion}

Typical diffusion models~\cite{zhang2023adding} rely on a sequence of noise-reduction steps based purely on learned data distributions without explicit physics-based guidance, which limits their effectiveness in applications requiring high precision. SpecDiffusion stands out from conventional diffusion models by incorporating speckle patterns as conditioning inputs, enhancing its capability for detailed and accurate image reconstruction in biomedical imaging. By utilizing physics-informed guidance in its denoising process, SpecDiffusion not only outperforms traditional models like U-Net in performance but also meets the specific demands of fiber endomicroscopic imaging more effectively. Furthermore, we have validated the diffusion-driven fiber endomicroscope's performance in cell segmentation, this capability can be further extended to other digital pathology applications, such as tumor boundary detection, tissue classification, and morphological analysis, with minimum invasiveness.

In the realm of generative models, SpecDiffusion presents notable advantages over established techniques such as generative adversarial network (GANs)~\cite{goodfellow2020generative} and vision transformers~\cite{han2022survey}. While GANs are adept at generating visually appealing images, they frequently encounter issues related to training stability and convergence, particularly in complex imaging tasks where the preservation of fine details is critical. Conversely, Vision Transformers, despite their efficacy in processing sequential data, demand substantial computational resources and often struggle to efficiently capture spatial hierarchies that are crucial for high-resolution imaging. SpecDiffusion employs an iterative noise reduction process that is both stable and robust, facilitating the high-resolution reconstruction of phases in a physics-informed manner. This approach not only enhances the model’s ability to handle detailed imaging tasks but also mitigates common problems associated with other generative models, such as training instability and excessive computational load. The integration of speckle patterns as a conditioning input further aids in guiding the denoising process, ensuring that SpecDiffusion maintains a high level of accuracy and detail essential for biomedical applications. Furthermore, the SpecDiffusion offers a resolution comparable to the iterative phase retrieval method ~\cite{sun2022quantitative}, while significantly reducing reconstruction time from 24 seconds per image to just 0.56 seconds. This substantial improvement in efficiency highlights SpecDiffusion's potential for real-time applications in clinical applications. Moreover, Lensless fiber endomicroscope offers substantial advantages for in vivo imaging, allowing for deep tissue visualization with minimal invasiveness. We have demonstrated that, combined with the high-resolution and accurate phase reconstructions provided by SpecDiffusion, this system could be further employed for real-time, label-free imaging of tissues. When integrated with advanced AI models like SAM, the system enables precise cell segmentation and tumor boundary detection, enhancing the accuracy of cancer diagnosis and tissue classification. The minimally invasive nature of the fiber endomicroscope, along with its potential for high-fidelity imaging, makes this approach particularly suited for real-time diagnostics, surgical guidance, and early detection of diseases in hard-to-reach areas, improving both patient outcomes and digital pathology workflows.

Despite these promising results, there are limitations to the current implementation of SpecDiffusion. While the model is trained on a large dataset of 100,000 images, its performance may still be limited when applied to tissue types or imaging conditions not represented in the dataset. Additionally, although the reconstruction speed is a significant improvement, further optimizations may be necessary for real-time clinical applications. Future work could address these limitations by expanding the dataset to include more diverse tissue types and imaging conditions, as well as optimizing the model for faster, real-time applications. Additionally, exploring the integration of SpecDiffusion with other imaging modalities could provide more comprehensive diagnostic capabilities. Overall, SpecDiffusion represents a significant step forward in quantitative phase reconstruction for lensless fiber endomicroscopy, offering a powerful tool for improving precision in medical diagnostics, particularly in cancer detection and digital pathology with minimal invasiveness.

\section*{Methods}

\label{sec:methods}

\subsection*{Experimental setup}
We implement the optical setup illustrated in Fig.~S$1$ to acquire the training images for the diffusion-driven fiber endomicroscope. In this setup, phase images are projected onto an SLM and imaged by the fiber endomicroscope, with the speckle patterns captured by the detection system. A detailed description of the optical setup is provided in supplementary information. During the experiment, the SLM and detection camera were triggered together to ensure the phase images and speckle patterns were captured at the same time, ensuring accurate alignment in the training dataset. The distance between the far-field image plane for capturing speckles and the MCF facet on the detection side was $0.5$ mm.

\subsection*{Training dataset}
Utilizing the optical setup, we construct a tailored dataset for QPI in lensless fiber endomicroscopy. The natural complex images from the ImageNet dataset~\cite{deng2009imagenet} are converted to grayscale images, and further transformed to phase images. These phase images are resized to a resolution of $980\times 980$ pixels and subsequently zero-padded to $1920 \times 1080$ pixels to accommodate the display resolution of the SLM. Computer-generated holograms are then calculated and displayed on the SLM to achieve a precise holographic representation of the ground truth image. Compared to MNIST and Fashion-MNIST datasets in our previous study~\cite{sun2024calibration}, images from ImageNet are more complex and detailed, thus providing a more robust benchmark for evaluating phase reconstruction capabilities in real-world scenarios. In the experiment, we captured $100,000$ paired phase and speckle images, allocating $90,000$ pairs for training, and the remaining $10,000$ pairs for testing.

\subsection*{SpecDiffusion model}
\label{sec:model}
Our proposed SpecDiffusion model employs a conditional diffusion pipeline, decomposing the phase reconstruction process into $T$ sequential image denoising steps. This method significantly mitigates the complexity of phase reconstruction for each individual step. As depicted in Fig.~\ref{fig:model}a, the SpecDiffusion training process adheres to a Markov Chain framework. In image denoising task $t$, it begins with diffusing target phase $\phi_0$ with Gaussian noise $\epsilon_t$, generating a pseudo-input to be denoised. This diffusion process is controlled by the diffusion coefficient $\overline{\alpha_t}$, as illustrated in Eq.1. Subsequently, this pseudo-input fuses with speckle $A$ to form a composite input, thereby incorporating speckle $A$ information to supervise the following image denoising process by SpecDiffusion. The goal of SpecDiffusion is to accurately estimate the added Gaussian noise $\epsilon_t$, compare its estimate with the actual noise and refine its parameters through back-propagation. Notably, as $t$ increases, the diffusion coefficient $\overline{\alpha_t}$ also increases, transforming the target phase towards approximate Gaussian noise. As the target phase information within the composite input is gradually erased in the diffusion process, SpecDiffusion must rely on the speckle pattern to predict the added Gaussian noise, and it also potentially alters its prediction goal from mere Gaussian noise estimation to a modification guiding the diffused input closer to the authentic phase image.

\begin{equation}
\phi_t = \sqrt{\overline{\alpha}_t} \phi_{t-1} + \sqrt{1-\overline{\alpha}_t}\epsilon_t.
\end{equation}

In the inference process, SpecDiffusion accomplishes phase reconstruction via $T$ sequential image denoising steps, methodically transforming Gaussian noise into the desired phase image. This process begins with a randomly generated initial phase $\phi_T'$, which is subsequently combined with the given speckle input $x$ to create a composite input for the SpecDiffusion model. Leveraging the speckle information $A$ within this composite input, SpecDiffusion estimates the noise $\epsilon_T'$ and eliminates it from $\phi_T'$, yielding a denoised phase $\phi_{T-1}'$. However, due to SpecDiffusion's inherent limitations, $\phi_{T-1}'$ typically deviates from the exact target phase. To refine the predicted phase image, $\phi_{T-1}'$ is re-noised with Gaussian noise $\epsilon_{T-1}'$ and then utilized as the initial phase for the subsequent image denoising step $T-1$. The initial phase for any given denoising step $t$ is obtained as depicted in Eq.2, maintaining the same diffusion ratio $\alpha_t$ as in the training process. Through $T$ denoising iterations, the phase image is progressively refined, leading to a final output accurately replicates the target phase.

\begin{equation}
\phi_{t-1}' = f_\theta(\sqrt{\overline{\alpha}_{t-1}}\phi_t' + \sqrt{1-\overline{\alpha}_{t-1}}\epsilon_{t-1}', A).
\end{equation}

\subsection*{Neural network training}
\label{sec:train}
SpecDiffusion is implemented using the Pytorch framework and Python $3.9.12$, and trained on a platform with GeForce RTX $3090$ GPU and Intel Xeon Platinum $8369$B CPU. Before training, speckle patterns and images are cropped to a size of $128\times128$. During training, we set the batch size to $16$ and learning rate to $10^{-4}$. L1 distance between the predicted noise and actual loss is adopted as the loss function. The diffusion ratio of SpecDiffusion is set to be $1e-6$ initially and linearly increases to $0.01$ over $2,000$ steps. Training proceeds for $400$ epochs with the Adam optimizer~\cite{kingma2014adam} and lasts 5.5 days for imageNet. During testing, SpecDiffusion is tested on a platform with NVIDIA Tesla A100 GPU and Intel(R) Xeon(R) Gold $6248$R CPU. The diffusion ratio of SpecDiffusion is set to be $1e-4$ initially and linearly increases to $0.09$ over $1,000$ steps.

\subsection*{Evaluation metrics}
\label{sec:metric}
In our experiment, we employ Mean Absolute Error (MAE), Peak Signal-to-Noise Ratio (PSNR), Structural Similarity Index Measure (SSIM), 2D Correlation coefficient as the evaluation metrics for reconstructed phase images. These metrics provide a quantitative measure of the reconstruction fidelity.

The mean absolute error (MAE) is a critical metric for assessing the accuracy of reconstructed images, particularly in the context of phase reconstruction. It measures the average of the absolute differences between the original phase image and the reconstructed phase image, across all pixels. A lower MAE value indicates a closer match between the reconstructed and original phases, signifying higher fidelity in the reconstruction process. This high fidelity is crucial for applications that rely on precise phase information, such as in the calculation of biophysical properties including refractive index, cell volume and dry mass. For original image $I$ and reconstructed image $I'$, the MAE is calculated as:
\begin{equation}
    \text{MAE}(I, I') = \frac{1}{mn}\sum\limits_{i=0}^{m-1} \sum\limits_{j=0}^{n-1} |I(i, j)-I'(i,j)|,
\end{equation}
where $m$ and $n$ indicate the row and column index of the image.

The peak signal-to-noise ratio (PSNR) is a widely utilized metric in the field of image processing to evaluate the quality of reconstructed images relative to their originals. It quantifies the ratio of the maximum possible power of the original image to the power of corrupting noise that affects the fidelity of the reconstructed image. Essentially, PSNR measures the level of distortion introduced into the original image upon reconstruction, which also reflects the amount of information derived from the speckle. A higher PSNR value typically indicates more information derived from the speckle, implying a higher quality of the reconstructed image. For original image $I$ and reconstructed image $I'$, the PSNR is calculated as:
\begin{equation}
\text{PSNR}(I, I') = 10\cdot \text{log}_{10}[\frac{\text{MAX}_I^2}{\text{MSE}(I, I')}],
\end{equation}
where $\text{MAX}_I$ is the max pixel value of image $I$ and $I'$, $\text{MSE}$ is the mean square error between image $I$ and $I'$, which is calculated as 

\begin{equation}
\text{MSE}(I, I') = \frac{1}{mn} \sum\limits_{i=0}^{m-1} \sum\limits_{j=0}^{n-1} [I(i,j) - I'(i,j)]^2,
\end{equation}
where $m$ and $n$ indicate the row and column index of the image.

The structural similarity of index measure (SSIM)~\cite{wang2004image} is an advanced metric designed to assess the perceptual quality of images. Unlike traditional metrics that primarily focus on pixel-level differences, SSIM evaluates the visual impact of three key components: luminance, contrast, and structure, which correspond to the perception of brightness, contrast, and patterns or textures in human vision, respectively. For original image $I$ and reconstructed image $I'$, the SSIM is calculated as 
\begin{equation}
\text{SSIM}(I, I') = \frac{(2\mu_I\mu_{I'}+c_1)(2\sigma_{II'}+c_2)}{(\mu_I^2+\mu_{I'}^2+c_1)(\sigma_I^2+\sigma_{I'}^2+c_2)},
\end{equation}
where $\mu$ indicates pixel sample mean, $\sigma$ indicates the variance, $c_1$ and $c_2$ are two variables to stabilize the division to avoid zero denominator.

The 2D correlation coefficient is a statistical measure that quantifies the degree of linear correlation between two images. It serves as an indicator of the overall fidelity of the reconstructed image by assessing how well the pixel intensity variations in the reconstructed image match those in the original image. For original image $I$ and reconstructed image $I'$, the 2D correlation coefficient is calculated as 
\begin{equation}
    \rho_{II'} = \frac{\text{Cov}(I, I')}{\sigma_I \sigma_{I'}},
\end{equation}
where $\text{Cov}(I, I')$ is the convariance between $I$ and $I'$, $\sigma_I$ and $\sigma_{I'}$ are the standard variance of $I$ and $I'$.

The PRSTD is employed to quantify the reconstruction bias in the reconstructed images. Typically, an image's background occupies a substantial and homogeneous area, making it easier to reconstruct accurately. Thus, data-driven models exhibit a reconstruction bias, favoring the background due to its simplicity. In contrast, the foreground, which contains more informative and complex features, is more challenging to be reconstructed and thus neglected. This bias between background and foreground reconstruction quality is captured by PRSTD. For original image $I$ and reconstructed image $I'$, the PRSTD is calculated as 
\begin{equation}
    \text{PRSTD}(I, I') = \sigma(|I-I'|),
\end{equation}
where $\sigma$ indicates standard variance.

For assessing the accuracy of cell segmentation on the reconstructed images, we adopted the intersection over union (IoU) metric. IoU quantifies the overlap between two segmentation maps, thereby providing an assessment of the similarity between the predicted segmentation result and the ground truth. This metric is particularly effective for comparing the precision of segmentation boundaries and the overall segmentation quality. For segmentation map $M$ of ground truth and $M'$ of reconstructed image, IoU is computed as 
\begin{equation}
    \text{IoU}(M, M') = \frac{|M\cap M'|}{|M\cup M'|},
\end{equation}
where $|M\cap M'|$ indicates the overlap area between $M$ and $M'$, $|M\cap M'|$ indicates the union area between $M$ and $M'$. 

\bibliography{sample}

\begin{thebibliography}{10}
\urlstyle{rm}
\expandafter\ifx\csname url\endcsname\relax
  \def\url#1{\texttt{#1}}\fi
\expandafter\ifx\csname urlprefix\endcsname\relax\def\urlprefix{URL }\fi
\expandafter\ifx\csname doiprefix\endcsname\relax\def\doiprefix{DOI: }\fi
\providecommand{\bibinfo}[2]{#2}
\providecommand{\eprint}[2][]{\url{#2}}

\bibitem{porat2016widefield}
\bibinfo{author}{Porat, A.} \emph{et~al.}
\newblock \bibinfo{journal}{\bibinfo{title}{Widefield lensless imaging through a fiber bundle via speckle correlations}}.
\newblock {\emph{\JournalTitle{Optics express}}} \textbf{\bibinfo{volume}{24}}, \bibinfo{pages}{16835--16855} (\bibinfo{year}{2016}).

\bibitem{borhani2018learning}
\bibinfo{author}{Borhani, N.}, \bibinfo{author}{Kakkava, E.}, \bibinfo{author}{Moser, C.} \& \bibinfo{author}{Psaltis, D.}
\newblock \bibinfo{journal}{\bibinfo{title}{Learning to see through multimode fibers}}.
\newblock {\emph{\JournalTitle{Optica}}} \textbf{\bibinfo{volume}{5}}, \bibinfo{pages}{960--966} (\bibinfo{year}{2018}).

\bibitem{kuschmierz2021ultra}
\bibinfo{author}{Kuschmierz, R.}, \bibinfo{author}{Scharf, E.}, \bibinfo{author}{Orteg{\'o}n-Gonz{\'a}lez, D.~F.}, \bibinfo{author}{Glosemeyer, T.} \& \bibinfo{author}{Czarske, J.~W.}
\newblock \bibinfo{journal}{\bibinfo{title}{Ultra-thin 3d lensless fiber endoscopy using diffractive optical elements and deep neural networks}}.
\newblock {\emph{\JournalTitle{Light: Advanced Manufacturing}}} \textbf{\bibinfo{volume}{2}}, \bibinfo{pages}{1--10} (\bibinfo{year}{2021}).

\bibitem{sun2024ai}
\bibinfo{author}{Sun, J.}, \bibinfo{author}{Yang, B.}, \bibinfo{author}{Koukourakis, N.}, \bibinfo{author}{Guck, J.} \& \bibinfo{author}{Czarske, J.~W.}
\newblock \bibinfo{journal}{\bibinfo{title}{Ai-driven projection tomography with multicore fibre-optic cell rotation}}.
\newblock {\emph{\JournalTitle{Nature Communications}}} \textbf{\bibinfo{volume}{15}}, \bibinfo{pages}{147} (\bibinfo{year}{2024}).

\bibitem{sun2022quantitative}
\bibinfo{author}{Sun, J.} \emph{et~al.}
\newblock \bibinfo{journal}{\bibinfo{title}{Quantitative phase imaging through an ultra-thin lensless fiber endoscope}}.
\newblock {\emph{\JournalTitle{Light: Science \& Applications}}} \textbf{\bibinfo{volume}{11}}, \bibinfo{pages}{204} (\bibinfo{year}{2022}).

\bibitem{badt2022real}
\bibinfo{author}{Badt, N.} \& \bibinfo{author}{Katz, O.}
\newblock \bibinfo{journal}{\bibinfo{title}{Real-time holographic lensless micro-endoscopy through flexible fibers via fiber bundle distal holography}}.
\newblock {\emph{\JournalTitle{Nature Communications}}} \textbf{\bibinfo{volume}{13}}, \bibinfo{pages}{6055} (\bibinfo{year}{2022}).

\bibitem{choi2022flexible}
\bibinfo{author}{Choi, W.} \emph{et~al.}
\newblock \bibinfo{journal}{\bibinfo{title}{Flexible-type ultrathin holographic endoscope for microscopic imaging of unstained biological tissues}}.
\newblock {\emph{\JournalTitle{Nature communications}}} \textbf{\bibinfo{volume}{13}}, \bibinfo{pages}{4469} (\bibinfo{year}{2022}).

\bibitem{wang2024resolution}
\bibinfo{author}{Wang, T.} \emph{et~al.}
\newblock \bibinfo{journal}{\bibinfo{title}{Resolution-enhanced multi-core fiber imaging learned on a digital twin for cancer diagnosis}}.
\newblock {\emph{\JournalTitle{Neurophotonics}}} \textbf{\bibinfo{volume}{11}}, \bibinfo{pages}{S11505--S11505} (\bibinfo{year}{2024}).

\bibitem{sun2024lensless}
\bibinfo{author}{Sun, J.}, \bibinfo{author}{Kuschmierz, R.}, \bibinfo{author}{Katz, O.}, \bibinfo{author}{Koukourakis, N.} \& \bibinfo{author}{Czarske, J.~W.}
\newblock \bibinfo{journal}{\bibinfo{title}{Lensless fiber endomicroscopy in biomedicine}}.
\newblock {\emph{\JournalTitle{PhotoniX}}} \textbf{\bibinfo{volume}{5}}, \bibinfo{pages}{18} (\bibinfo{year}{2024}).

\bibitem{li2021memory}
\bibinfo{author}{Li, S.}, \bibinfo{author}{Horsley, S.~A.}, \bibinfo{author}{Tyc, T.}, \bibinfo{author}{{\v{C}}i{\v{z}}m{\'a}r, T.} \& \bibinfo{author}{Phillips, D.~B.}
\newblock \bibinfo{journal}{\bibinfo{title}{Memory effect assisted imaging through multimode optical fibres}}.
\newblock {\emph{\JournalTitle{Nature Communications}}} \textbf{\bibinfo{volume}{12}}, \bibinfo{pages}{1--13} (\bibinfo{year}{2021}).

\bibitem{koukourakis2022investigation}
\bibinfo{author}{Koukourakis, N.}, \bibinfo{author}{Wagner, F.}, \bibinfo{author}{Rothe, S.}, \bibinfo{author}{Karl, M.~O.} \& \bibinfo{author}{Czarske, J.~W.}
\newblock \bibinfo{journal}{\bibinfo{title}{Investigation of human organoid retina with digital holographic transmission matrix measurements}}.
\newblock {\emph{\JournalTitle{Light: Advanced Manufacturing}}} \textbf{\bibinfo{volume}{3}}, \bibinfo{pages}{211--225} (\bibinfo{year}{2022}).

\bibitem{lich2024single}
\bibinfo{author}{Lich, J.}, \bibinfo{author}{Glosemeyer, T.}, \bibinfo{author}{Czarske, J.} \& \bibinfo{author}{Kuschmierz, R.}
\newblock \bibinfo{journal}{\bibinfo{title}{Single-shot 3d incoherent imaging with diffuser endoscopy}}.
\newblock {\emph{\JournalTitle{Light: Advanced Manufacturing}}} \textbf{\bibinfo{volume}{5}}, \bibinfo{pages}{1--11} (\bibinfo{year}{2024}).

\bibitem{schurmann2018three}
\bibinfo{author}{Sch{\"u}rmann, M.} \emph{et~al.}
\newblock \bibinfo{journal}{\bibinfo{title}{Three-dimensional correlative single-cell imaging utilizing fluorescence and refractive index tomography}}.
\newblock {\emph{\JournalTitle{Journal of biophotonics}}} \textbf{\bibinfo{volume}{11}}, \bibinfo{pages}{e201700145} (\bibinfo{year}{2018}).

\bibitem{sun2021complex}
\bibinfo{author}{Sun, J.}, \bibinfo{author}{Koukourakis, N.} \& \bibinfo{author}{Czarske, J.~W.}
\newblock \bibinfo{journal}{\bibinfo{title}{Complex wavefront shaping through a multi-core fiber}}.
\newblock {\emph{\JournalTitle{Applied Sciences}}} \textbf{\bibinfo{volume}{11}}, \bibinfo{pages}{3949} (\bibinfo{year}{2021}).

\bibitem{flusberg2005fiber}
\bibinfo{author}{Flusberg, B.~A.} \emph{et~al.}
\newblock \bibinfo{journal}{\bibinfo{title}{Fiber-optic fluorescence imaging}}.
\newblock {\emph{\JournalTitle{Nature methods}}} \textbf{\bibinfo{volume}{2}}, \bibinfo{pages}{941--950} (\bibinfo{year}{2005}).

\bibitem{szabo2014spatially}
\bibinfo{author}{Szabo, V.}, \bibinfo{author}{Ventalon, C.}, \bibinfo{author}{De~Sars, V.}, \bibinfo{author}{Bradley, J.} \& \bibinfo{author}{Emiliani, V.}
\newblock \bibinfo{journal}{\bibinfo{title}{Spatially selective holographic photoactivation and functional fluorescence imaging in freely behaving mice with a fiberscope}}.
\newblock {\emph{\JournalTitle{Neuron}}} \textbf{\bibinfo{volume}{84}}, \bibinfo{pages}{1157--1169} (\bibinfo{year}{2014}).

\bibitem{haufe2017transmission}
\bibinfo{author}{Haufe, D.}, \bibinfo{author}{Koukourakis, N.}, \bibinfo{author}{B{\"u}ttner, L.} \& \bibinfo{author}{Czarske, J.~W.}
\newblock \bibinfo{journal}{\bibinfo{title}{Transmission of multiple signals through an optical fiber using wavefront shaping}}.
\newblock {\emph{\JournalTitle{Journal of Visualized Experiments: JoVE}}}  (\bibinfo{year}{2017}).

\bibitem{rothe2019transmission}
\bibinfo{author}{Rothe, S.}, \bibinfo{author}{Radner, H.}, \bibinfo{author}{Koukourakis, N.} \& \bibinfo{author}{Czarske, J.~W.}
\newblock \bibinfo{journal}{\bibinfo{title}{Transmission matrix measurement of multimode optical fibers by mode-selective excitation using one spatial light modulator}}.
\newblock {\emph{\JournalTitle{Applied Sciences}}} \textbf{\bibinfo{volume}{9}}, \bibinfo{pages}{195} (\bibinfo{year}{2019}).

\bibitem{abraham2023label}
\bibinfo{author}{Abraham, T.~M.} \emph{et~al.}
\newblock \bibinfo{journal}{\bibinfo{title}{Label-and slide-free tissue histology using 3d epi-mode quantitative phase imaging and virtual hematoxylin and eosin staining}}.
\newblock {\emph{\JournalTitle{Optica}}} \textbf{\bibinfo{volume}{10}}, \bibinfo{pages}{1605--1618} (\bibinfo{year}{2023}).

\bibitem{park2023artificial}
\bibinfo{author}{Park, J.} \emph{et~al.}
\newblock \bibinfo{journal}{\bibinfo{title}{Artificial intelligence-enabled quantitative phase imaging methods for life sciences}}.
\newblock {\emph{\JournalTitle{Nature Methods}}} \textbf{\bibinfo{volume}{20}}, \bibinfo{pages}{1645--1660} (\bibinfo{year}{2023}).

\bibitem{pillar2024virtual}
\bibinfo{author}{Pillar, N.}, \bibinfo{author}{Li, Y.}, \bibinfo{author}{Zhang, Y.} \& \bibinfo{author}{Ozcan, A.}
\newblock \bibinfo{journal}{\bibinfo{title}{Virtual staining of non-fixed tissue histology}}.
\newblock {\emph{\JournalTitle{Modern Pathology}}} \bibinfo{pages}{100444} (\bibinfo{year}{2024}).

\bibitem{rothe2021benchmarking}
\bibinfo{author}{Rothe, S.} \emph{et~al.}
\newblock \bibinfo{journal}{\bibinfo{title}{Benchmarking analysis of computer generated holograms for complex wavefront shaping using pixelated phase modulators}}.
\newblock {\emph{\JournalTitle{Optics Express}}} \textbf{\bibinfo{volume}{29}}, \bibinfo{pages}{37602--37616} (\bibinfo{year}{2021}).

\bibitem{wang2011tissue}
\bibinfo{author}{Wang, Z.}, \bibinfo{author}{Tangella, K.}, \bibinfo{author}{Balla, A.} \& \bibinfo{author}{Popescu, G.}
\newblock \bibinfo{journal}{\bibinfo{title}{Tissue refractive index as marker of disease}}.
\newblock {\emph{\JournalTitle{Journal of biomedical optics}}} \textbf{\bibinfo{volume}{16}}, \bibinfo{pages}{116017--116017} (\bibinfo{year}{2011}).

\bibitem{Liu2016}
\bibinfo{author}{Liu, P.~Y.} \emph{et~al.}
\newblock \bibinfo{journal}{\bibinfo{title}{{Cell refractive index for cell biology and disease diagnosis: Past, present and future}}}.
\newblock {\emph{\JournalTitle{Lab on a Chip}}} \textbf{\bibinfo{volume}{16}}, \bibinfo{pages}{634--644}, \doiprefix\url{10.1039/c5lc01445j} (\bibinfo{year}{2016}).

\bibitem{park2018quantitative}
\bibinfo{author}{Park, Y.}, \bibinfo{author}{Depeursinge, C.} \& \bibinfo{author}{Popescu, G.}
\newblock \bibinfo{journal}{\bibinfo{title}{Quantitative phase imaging in biomedicine}}.
\newblock {\emph{\JournalTitle{Nature photonics}}} \textbf{\bibinfo{volume}{12}}, \bibinfo{pages}{578--589} (\bibinfo{year}{2018}).

\bibitem{schurmann2016cell}
\bibinfo{author}{Sch{\"u}rmann, M.}, \bibinfo{author}{Scholze, J.}, \bibinfo{author}{M{\"u}ller, P.}, \bibinfo{author}{Guck, J.} \& \bibinfo{author}{Chan, C.~J.}
\newblock \bibinfo{journal}{\bibinfo{title}{Cell nuclei have lower refractive index and mass density than cytoplasm}}.
\newblock {\emph{\JournalTitle{Journal of biophotonics}}} \textbf{\bibinfo{volume}{9}}, \bibinfo{pages}{1068--1076} (\bibinfo{year}{2016}).

\bibitem{aknoun2015living}
\bibinfo{author}{Aknoun, S.} \emph{et~al.}
\newblock \bibinfo{journal}{\bibinfo{title}{Living cell dry mass measurement using quantitative phase imaging with quadriwave lateral shearing interferometry: an accuracy and sensitivity discussion}}.
\newblock {\emph{\JournalTitle{Journal of biomedical optics}}} \textbf{\bibinfo{volume}{20}}, \bibinfo{pages}{126009--126009} (\bibinfo{year}{2015}).

\bibitem{parvin2021differential}
\bibinfo{author}{Parvin, T.}, \bibinfo{author}{Ahmed, K.}, \bibinfo{author}{Alatwi, A.~M.} \& \bibinfo{author}{Rashed, A. N.~Z.}
\newblock \bibinfo{journal}{\bibinfo{title}{Differential optical absorption spectroscopy-based refractive index sensor for cancer cell detection}}.
\newblock {\emph{\JournalTitle{Optical Review}}} \textbf{\bibinfo{volume}{28}}, \bibinfo{pages}{134--143} (\bibinfo{year}{2021}).

\bibitem{kuschmierz2018}
\bibinfo{author}{Kuschmierz, R.}, \bibinfo{author}{Scharf, E.}, \bibinfo{author}{Koukourakis, N.} \& \bibinfo{author}{Czarske, J.~W.}
\newblock \bibinfo{journal}{\bibinfo{title}{{Self-calibration of lensless holographic endoscope using programmable guide stars}}}.
\newblock {\emph{\JournalTitle{Optics Letters}}} \textbf{\bibinfo{volume}{43}}, \bibinfo{pages}{2997}, \doiprefix\url{10.1364/ol.43.002997} (\bibinfo{year}{2018}).

\bibitem{sun2023compressive}
\bibinfo{author}{Sun, J.} \& \bibinfo{author}{Czarske, J.~W.}
\newblock \bibinfo{journal}{\bibinfo{title}{Compressive holographic sensing simplifies quantitative phase imaging}}.
\newblock {\emph{\JournalTitle{Light: Science \& Applications}}} \textbf{\bibinfo{volume}{12}}, \bibinfo{pages}{121} (\bibinfo{year}{2023}).

\bibitem{wang2020phase}
\bibinfo{author}{Wang, F.} \emph{et~al.}
\newblock \bibinfo{journal}{\bibinfo{title}{Phase imaging with an untrained neural network}}.
\newblock {\emph{\JournalTitle{Light: Science \& Applications}}} \textbf{\bibinfo{volume}{9}}, \bibinfo{pages}{77} (\bibinfo{year}{2020}).

\bibitem{bostan2020deep}
\bibinfo{author}{Bostan, E.}, \bibinfo{author}{Heckel, R.}, \bibinfo{author}{Chen, M.}, \bibinfo{author}{Kellman, M.} \& \bibinfo{author}{Waller, L.}
\newblock \bibinfo{journal}{\bibinfo{title}{Deep phase decoder: self-calibrating phase microscopy with an untrained deep neural network}}.
\newblock {\emph{\JournalTitle{Optica}}} \textbf{\bibinfo{volume}{7}}, \bibinfo{pages}{559--562} (\bibinfo{year}{2020}).

\bibitem{zhou2024deep}
\bibinfo{author}{Zhou, J.} \emph{et~al.}
\newblock \bibinfo{journal}{\bibinfo{title}{Deep learning-enabled pixel-super-resolved quantitative phase microscopy from single-shot aliased intensity measurement}}.
\newblock {\emph{\JournalTitle{Laser \& Photonics Reviews}}} \textbf{\bibinfo{volume}{18}}, \bibinfo{pages}{2300488} (\bibinfo{year}{2024}).

\bibitem{liu2024learning}
\bibinfo{author}{Liu, H.} \emph{et~al.}
\newblock \bibinfo{journal}{\bibinfo{title}{Learning-based real-time imaging through dynamic scattering media}}.
\newblock {\emph{\JournalTitle{Light: Science \& Applications}}} \textbf{\bibinfo{volume}{13}}, \bibinfo{pages}{194} (\bibinfo{year}{2024}).

\bibitem{wu2021high}
\bibinfo{author}{Wu, J.}, \bibinfo{author}{Liu, K.}, \bibinfo{author}{Sui, X.} \& \bibinfo{author}{Cao, L.}
\newblock \bibinfo{journal}{\bibinfo{title}{High-speed computer-generated holography using an autoencoder-based deep neural network}}.
\newblock {\emph{\JournalTitle{Optics Letters}}} \textbf{\bibinfo{volume}{46}}, \bibinfo{pages}{2908--2911} (\bibinfo{year}{2021}).

\bibitem{zhang2022learning}
\bibinfo{author}{Zhang, Q.}, \bibinfo{author}{Rothe, S.}, \bibinfo{author}{Koukourakis, N.} \& \bibinfo{author}{Czarske, J.}
\newblock \bibinfo{journal}{\bibinfo{title}{Learning the matrix of few-mode fibers for high-fidelity spatial mode transmission}}.
\newblock {\emph{\JournalTitle{APL Photonics}}} \textbf{\bibinfo{volume}{7}} (\bibinfo{year}{2022}).

\bibitem{rahmani2018multimode}
\bibinfo{author}{Rahmani, B.}, \bibinfo{author}{Loterie, D.}, \bibinfo{author}{Konstantinou, G.}, \bibinfo{author}{Psaltis, D.} \& \bibinfo{author}{Moser, C.}
\newblock \bibinfo{journal}{\bibinfo{title}{Multimode optical fiber transmission with a deep learning network}}.
\newblock {\emph{\JournalTitle{Light: Science \& Applications}}} \textbf{\bibinfo{volume}{7}}, \bibinfo{pages}{1--11} (\bibinfo{year}{2018}).

\bibitem{chen2023deep}
\bibinfo{author}{Chen, Y.}, \bibinfo{author}{Song, B.}, \bibinfo{author}{Wu, J.}, \bibinfo{author}{Lin, W.} \& \bibinfo{author}{Huang, W.}
\newblock \bibinfo{journal}{\bibinfo{title}{Deep learning for efficiently imaging through the localized speckle field of a multimode fiber}}.
\newblock {\emph{\JournalTitle{Applied Optics}}} \textbf{\bibinfo{volume}{62}}, \bibinfo{pages}{266--274} (\bibinfo{year}{2023}).

\bibitem{lecun1998gradient}
\bibinfo{author}{LeCun, Y.}, \bibinfo{author}{Bottou, L.}, \bibinfo{author}{Bengio, Y.} \& \bibinfo{author}{Haffner, P.}
\newblock \bibinfo{journal}{\bibinfo{title}{Gradient-based learning applied to document recognition}}.
\newblock {\emph{\JournalTitle{Proceedings of the IEEE}}} \textbf{\bibinfo{volume}{86}}, \bibinfo{pages}{2278--2324} (\bibinfo{year}{1998}).

\bibitem{sun2024calibration}
\bibinfo{author}{Sun, J.} \emph{et~al.}
\newblock \bibinfo{journal}{\bibinfo{title}{Calibration-free quantitative phase imaging in multi-core fiber endoscopes using end-to-end deep learning}}.
\newblock {\emph{\JournalTitle{Optics Letters}}} \textbf{\bibinfo{volume}{49}}, \bibinfo{pages}{342--345} (\bibinfo{year}{2024}).

\bibitem{deng2012mnist}
\bibinfo{author}{Deng, L.}
\newblock \bibinfo{journal}{\bibinfo{title}{The mnist database of handwritten digit images for machine learning research [best of the web]}}.
\newblock {\emph{\JournalTitle{IEEE Signal Processing Magazine}}} \textbf{\bibinfo{volume}{29}}, \bibinfo{pages}{141--142} (\bibinfo{year}{2012}).

\bibitem{xiao2017fashion}
\bibinfo{author}{Xiao, H.}, \bibinfo{author}{Rasul, K.} \& \bibinfo{author}{Vollgraf, R.}
\newblock \bibinfo{journal}{\bibinfo{title}{Fashion-mnist: a novel image dataset for benchmarking machine learning algorithms}}.
\newblock {\emph{\JournalTitle{arXiv preprint arXiv:1708.07747}}}  (\bibinfo{year}{2017}).

\bibitem{ronneberger2015u}
\bibinfo{author}{Ronneberger, O.}, \bibinfo{author}{Fischer, P.} \& \bibinfo{author}{Brox, T.}
\newblock \bibinfo{title}{U-net: Convolutional networks for biomedical image segmentation}.
\newblock In \emph{\bibinfo{booktitle}{International Conference on Medical image computing and computer-assisted intervention}}, \bibinfo{pages}{234--241} (\bibinfo{organization}{Springer}, \bibinfo{year}{2015}).

\bibitem{song2019generative}
\bibinfo{author}{Song, Y.} \& \bibinfo{author}{Ermon, S.}
\newblock \bibinfo{journal}{\bibinfo{title}{Generative modeling by estimating gradients of the data distribution}}.
\newblock {\emph{\JournalTitle{Advances in neural information processing systems}}} \textbf{\bibinfo{volume}{32}} (\bibinfo{year}{2019}).

\bibitem{ho2020denoising}
\bibinfo{author}{Ho, J.}, \bibinfo{author}{Jain, A.} \& \bibinfo{author}{Abbeel, P.}
\newblock \bibinfo{journal}{\bibinfo{title}{Denoising diffusion probabilistic models}}.
\newblock {\emph{\JournalTitle{Advances in neural information processing systems}}} \textbf{\bibinfo{volume}{33}}, \bibinfo{pages}{6840--6851} (\bibinfo{year}{2020}).

\bibitem{rombach2022high}
\bibinfo{author}{Rombach, R.}, \bibinfo{author}{Blattmann, A.}, \bibinfo{author}{Lorenz, D.}, \bibinfo{author}{Esser, P.} \& \bibinfo{author}{Ommer, B.}
\newblock \bibinfo{title}{High-resolution image synthesis with latent diffusion models}.
\newblock In \emph{\bibinfo{booktitle}{Proceedings of the IEEE/CVF conference on computer vision and pattern recognition}}, \bibinfo{pages}{10684--10695} (\bibinfo{year}{2022}).

\bibitem{igashov2024equivariant}
\bibinfo{author}{Igashov, I.} \emph{et~al.}
\newblock \bibinfo{journal}{\bibinfo{title}{Equivariant 3d-conditional diffusion model for molecular linker design}}.
\newblock {\emph{\JournalTitle{Nature Machine Intelligence}}} \bibinfo{pages}{1--11} (\bibinfo{year}{2024}).

\bibitem{guo2024diffusion}
\bibinfo{author}{Guo, Z.} \emph{et~al.}
\newblock \bibinfo{journal}{\bibinfo{title}{Diffusion models in bioinformatics and computational biology}}.
\newblock {\emph{\JournalTitle{Nature reviews bioengineering}}} \textbf{\bibinfo{volume}{2}}, \bibinfo{pages}{136--154} (\bibinfo{year}{2024}).

\bibitem{kreis2022latent}
\bibinfo{author}{Kreis, K.}, \bibinfo{author}{Dockhorn, T.}, \bibinfo{author}{Li, Z.} \& \bibinfo{author}{Zhong, E.}
\newblock \bibinfo{journal}{\bibinfo{title}{Latent space diffusion models of cryo-em structures}}.
\newblock {\emph{\JournalTitle{arXiv preprint arXiv:2211.14169}}}  (\bibinfo{year}{2022}).

\bibitem{waibel2023diffusion}
\bibinfo{author}{Waibel, D.~J.}, \bibinfo{author}{R{\"o}ell, E.}, \bibinfo{author}{Rieck, B.}, \bibinfo{author}{Giryes, R.} \& \bibinfo{author}{Marr, C.}
\newblock \bibinfo{title}{A diffusion model predicts 3d shapes from 2d microscopy images}.
\newblock In \emph{\bibinfo{booktitle}{2023 IEEE 20th International Symposium on Biomedical Imaging (ISBI)}}, \bibinfo{pages}{1--5} (\bibinfo{organization}{IEEE}, \bibinfo{year}{2023}).

\bibitem{wang2024conditional}
\bibinfo{author}{Wang, X.}, \bibinfo{author}{Yue, Q.} \& \bibinfo{author}{Liu, X.}
\newblock \bibinfo{journal}{\bibinfo{title}{Conditional diffusion model-based generation of speckle patterns for digital image correlation}}.
\newblock {\emph{\JournalTitle{Optics and Lasers in Engineering}}} \textbf{\bibinfo{volume}{175}}, \bibinfo{pages}{107997} (\bibinfo{year}{2024}).

\bibitem{wang2024decoding}
\bibinfo{author}{Wang, T.} \emph{et~al.}
\newblock \bibinfo{journal}{\bibinfo{title}{Decoding wavelengths from compressed speckle patterns with deep learning}}.
\newblock {\emph{\JournalTitle{Optics and Lasers in Engineering}}} \textbf{\bibinfo{volume}{180}}, \bibinfo{pages}{108268} (\bibinfo{year}{2024}).

\bibitem{zhang2024single}
\bibinfo{author}{Zhang, Y.}, \bibinfo{author}{Liu, X.} \& \bibinfo{author}{Lam, E.~Y.}
\newblock \bibinfo{journal}{\bibinfo{title}{Single-shot inline holography using a physics-aware diffusion model}}.
\newblock {\emph{\JournalTitle{Optics Express}}} \textbf{\bibinfo{volume}{32}}, \bibinfo{pages}{10444--10460} (\bibinfo{year}{2024}).

\bibitem{gerchberg1972holography}
\bibinfo{author}{Gerchberg, R.}
\newblock \bibinfo{journal}{\bibinfo{title}{Holography without fringes in the electron microscope}}.
\newblock {\emph{\JournalTitle{Nature}}} \textbf{\bibinfo{volume}{240}}, \bibinfo{pages}{404--406} (\bibinfo{year}{1972}).

\bibitem{deng2009imagenet}
\bibinfo{author}{Deng, J.} \emph{et~al.}
\newblock \bibinfo{title}{Imagenet: A large-scale hierarchical image database}.
\newblock In \emph{\bibinfo{booktitle}{2009 IEEE conference on computer vision and pattern recognition}}, \bibinfo{pages}{248--255} (\bibinfo{organization}{Ieee}, \bibinfo{year}{2009}).

\bibitem{wang2004image}
\bibinfo{author}{Wang, Z.}, \bibinfo{author}{Bovik, A.~C.}, \bibinfo{author}{Sheikh, H.~R.} \& \bibinfo{author}{Simoncelli, E.~P.}
\newblock \bibinfo{journal}{\bibinfo{title}{Image quality assessment: from error visibility to structural similarity}}.
\newblock {\emph{\JournalTitle{IEEE transactions on image processing}}} \textbf{\bibinfo{volume}{13}}, \bibinfo{pages}{600--612} (\bibinfo{year}{2004}).

\bibitem{durall2020watch}
\bibinfo{author}{Durall, R.}, \bibinfo{author}{Keuper, M.} \& \bibinfo{author}{Keuper, J.}
\newblock \bibinfo{title}{Watch your up-convolution: Cnn based generative deep neural networks are failing to reproduce spectral distributions}.
\newblock In \emph{\bibinfo{booktitle}{Proceedings of the IEEE/CVF conference on computer vision and pattern recognition}}, \bibinfo{pages}{7890--7899} (\bibinfo{year}{2020}).

\bibitem{kirillov2023segment}
\bibinfo{author}{Kirillov, A.} \emph{et~al.}
\newblock \bibinfo{title}{Segment anything}.
\newblock In \emph{\bibinfo{booktitle}{Proceedings of the IEEE/CVF International Conference on Computer Vision}}, \bibinfo{pages}{4015--4026} (\bibinfo{year}{2023}).

\bibitem{zhang2023adding}
\bibinfo{author}{Zhang, L.}, \bibinfo{author}{Rao, A.} \& \bibinfo{author}{Agrawala, M.}
\newblock \bibinfo{title}{Adding conditional control to text-to-image diffusion models}.
\newblock In \emph{\bibinfo{booktitle}{Proceedings of the IEEE/CVF International Conference on Computer Vision}}, \bibinfo{pages}{3836--3847} (\bibinfo{year}{2023}).

\bibitem{goodfellow2020generative}
\bibinfo{author}{Goodfellow, I.} \emph{et~al.}
\newblock \bibinfo{journal}{\bibinfo{title}{Generative adversarial networks}}.
\newblock {\emph{\JournalTitle{Communications of the ACM}}} \textbf{\bibinfo{volume}{63}}, \bibinfo{pages}{139--144} (\bibinfo{year}{2020}).

\bibitem{han2022survey}
\bibinfo{author}{Han, K.} \emph{et~al.}
\newblock \bibinfo{journal}{\bibinfo{title}{A survey on vision transformer}}.
\newblock {\emph{\JournalTitle{IEEE transactions on pattern analysis and machine intelligence}}} \textbf{\bibinfo{volume}{45}}, \bibinfo{pages}{87--110} (\bibinfo{year}{2022}).

\bibitem{kingma2014adam}
\bibinfo{author}{Kingma, D.~P.} \& \bibinfo{author}{Ba, J.}
\newblock \bibinfo{journal}{\bibinfo{title}{Adam: A method for stochastic optimization}}.
\newblock {\emph{\JournalTitle{arXiv preprint arXiv:1412.6980}}}  (\bibinfo{year}{2014}).

\bibitem{chen2024}
\bibinfo{author}{Chen, Z.} \emph{et~al.}
\newblock \bibinfo{title}{Specdiffusion for lensless fiber imaging}.
\newblock \bibinfo{howpublished}{\url{https://github.com/windbro98/SpecDiffusion-for-lensless-fiber-imaging}} (\bibinfo{year}{2024}).

\end{thebibliography}



\section*{Acknowledgements}
This work is supported by the Shanghai AI Laboratory, National Key R\&D Program of China (2022ZD0160102), the National Natural Science Foundation of China (62376222), and Young Elite Scientists Sponsorship Program by CAST (2023QNRC001). The authors would like to thank the valuable support from Dr. Nektarios Koukourakis.

\section*{Author contributions statement}
Z.C., J.S. and B.Z. conceived the experiments, Z.C. and J.S. conducted the experiments, Z.C. and X.Y. analysed the results, Z.C., J.S. wrote the manuscript, all authors reviewed the manuscript.

\section*{Disclosures}
The authors declare no conflicts of interest.

\section*{Data Availability Statement}
The SpecDiffusion model and the source code are publicly available on GitHub~\cite{chen2024}.

\section*{Supplemental document}
See Supplement information and supplementary video for supporting content.






\end{document}